\documentclass[twocolumn]{aastex6}
\usepackage[T1]{fontenc}
\usepackage{graphicx}
\usepackage{amsmath}
\usepackage{amsfonts}
\usepackage{amssymb}
\usepackage{bm}
\usepackage{natbib}
\usepackage{textcomp}
\newcommand{\sunrise}{\textsc{Sunrise}}

\begin{document}                 
\title{Solar Coronal Loops Associated with Small-scale Mixed Polarity Surface Magnetic Fields}

\author{\textsc{L. P. Chitta,$^{1}$ H. Peter,$^{1}$ 
S.~K.~Solanki,$^{1,2}$
P.~Barthol,$^{1}$
A.~Gandorfer,$^{1}$
L.~Gizon,$^{1,3}$
J.~Hirzberger,$^{1}$
T.~L.~Riethm\"uller,$^{1}$
M.~van~Noort,$^{1}$
J.~Blanco~Rodr\'{\i}guez,$^{4}$
J.~C.~Del~Toro~Iniesta,$^{5}$
D.~Orozco~Su\'arez,$^{5}$
W.~Schmidt,$^{6}$
V.~Mart\'{\i}nez Pillet,$^{7}$
\& M.~Kn\"olker$^{8}$}}
\affil{
$^{1}$Max-Planck-Institut f\"ur Sonnensystemforschung, Justus-von-Liebig-Weg 3, 37077 G\"ottingen, Germany; \href{mailto:chitta@mps.mpg.de}{chitta@mps.mpg.de}\\
$^{2}$School of Space Research, Kyung Hee University, Yongin, Gyeonggi, 446-701, Republic of Korea\\
$^{3}$Institut f\"ur Astrophysik, Georg-August-Universit\"at G\"ottingen, Friedrich-Hund-Platz 1, 37077 G\"ottingen, Germany\\
$^{4}$Grupo de Astronom\'{\i}a y Ciencias del Espacio, Universidad de Valencia, 46980 Paterna, Valencia, Spain\\
$^{5}$Instituto de Astrof\'{\i}sica de Andaluc\'{\i}a (CSIC), Apartado de Correos 3004, 18080 Granada, Spain\\
$^{6}$Kiepenheuer-Institut f\"ur Sonnenphysik, Sch\"oneckstr. 6, 79104 Freiburg, Germany\\
$^{7}$National Solar Observatory, 3665 Discovery Drive, Boulder, CO 80303, USA\\
$^{8}$High Altitude Observatory, National Center for Atmospheric Research,\footnote{The National Center for Atmospheric Research is sponsored by the National Science Foundation.} P.O. Box 3000, Boulder, CO 80307-3000, USA\\
}


\begin{abstract}
How and where are coronal loops rooted in the solar lower atmosphere? The details of the magnetic environment and its evolution at the footpoints of coronal loops are crucial to understanding the processes of mass and energy supply to the solar corona. To address the above question, we use high-resolution line-of-sight magnetic field data from the Imaging Magnetograph eXperiment instrument on the \sunrise{} balloon-borne observatory and coronal observations from the Atmospheric Imaging Assembly onboard the \textit{Solar Dynamics Observatory} of an emerging active region. We find that the coronal loops are often rooted at the locations with minor small-scale but persistent opposite-polarity magnetic elements very close to the larger dominant polarity. These opposite-polarity small-scale elements continually interact with the dominant polarity underlying the coronal loop through flux cancellation. At these locations we detect small inverse Y-shaped jets in chromospheric Ca\,{\sc ii}\,H images obtained from the \sunrise{} Filter Imager during the flux cancellation. Our results indicate that magnetic flux cancellation and reconnection at the base of coronal loops due to mixed polarity fields might be a crucial feature for the supply of mass and energy into the corona.
\end{abstract}


\section{Introduction}\label{sec:intro}   

The upper atmosphere of the Sun is dominated by the radiation from hot plasma appearing in the form of coronal loops. This emission in X-rays and the extreme UV (EUV) is aligned with the magnetic field and thus outlines the magnetic structure of the upper atmosphere. In the traditional picture, the magnetic field of a coronal loop is rooted at two footpoints with opposite magnetic polarities, thus connecting a north with a south magnetic polarity \citep{2005psci.book.....A}.

In this classic view, each footpoint of the coronal loop would be rooted in the photosphere in a unipolar magnetic region. In this work, we define unipolar magnetic regions as magnetic structures within which magnetic elements have the same polarity. For example, a typical spatially resolved unipolar fluxtube with a magnetic field strength of about 1000\,G has a size on the order of 100\,km diameter \citep[e.g.][see \citealt{1993SSRv...63....1S} for an extensive review on small-scale solar magnetic fields]{2010ApJ...723L.164L}. Based on the simple argument of magnetic flux conservation, an expansion of this kilogauss magnetic fluxtube into the corona to a diameter of a few megameters \citep[e.g.,][]{2011ApJ...732...81A} would imply a field strength of  some 10\,G at the loop apex. This is also consistent with coronal magnetic fields deduced from coronal seismology \citep[e.g.\ see the discussion in Sect.\,2 of][]{2015A&A...584A..68P}. Following this consideration, a coronal loop could indeed have a direct smooth connection down to the photosphere.

Such a direct magnetic connection is an (implicit) feature of any 1D coronal loop model \citep[for a review see, e.g.,][]{2014LRSP...11....4R}. It is also a feature of wave heating models of coronal loops \citep[e.g.][]{2011ApJ...736....3V} and models in which the braiding of magnetic field lines \citep[e.g.][]{1996JGR...10113445G} causes nanoflares \citep{1988ApJ...330..474P} in order to heat the corona. All of these models assume that there is a single magnetic polarity, i.e.\ a unipolar magnetic patch, at each of the two footpoints. This is true also for more complex 3D MHD active region coronal models, whether for a stable active region \citep{2011A&A...530A.112B,2012A&A...548A...1P} or a newly emerging active region \citep{2015NatPh..11..492C}.

\begin{figure*}[t]
\begin{center}
\includegraphics[width=\textwidth]{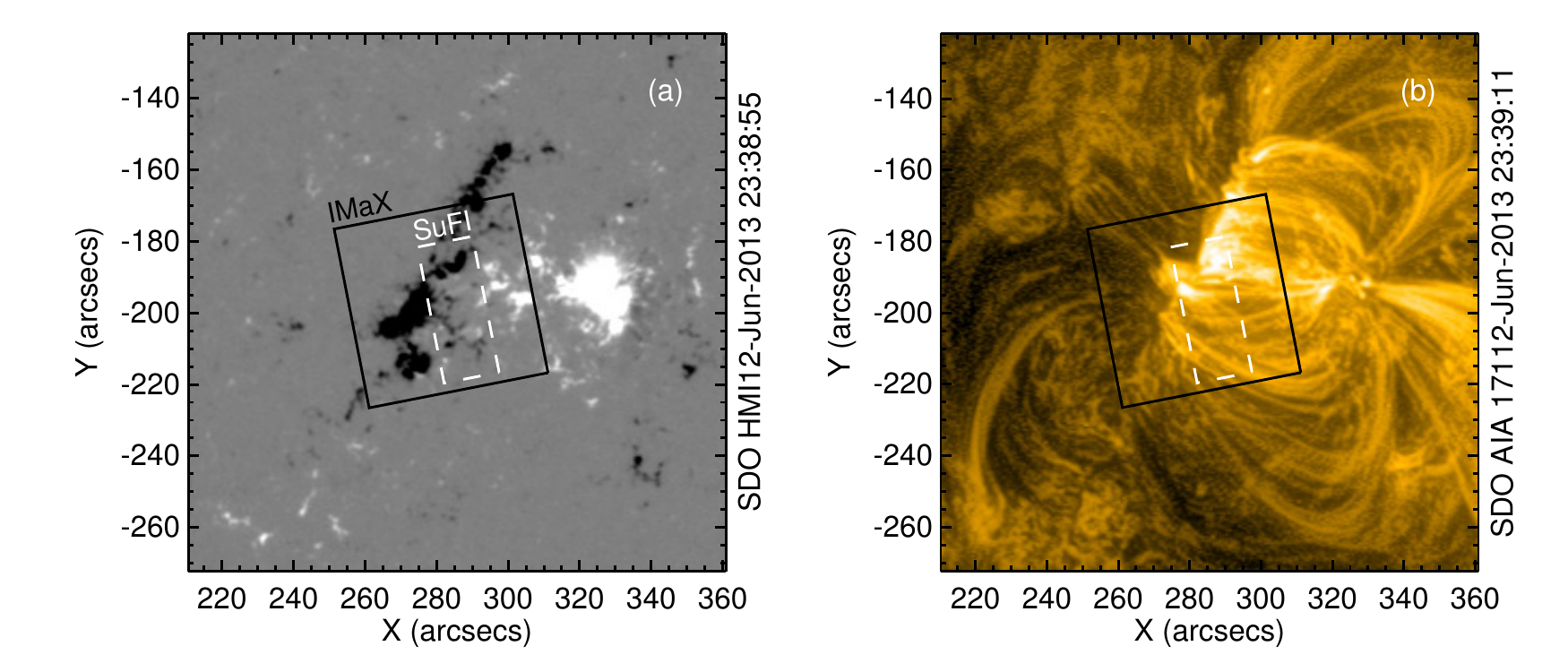}
\caption{Contextual maps of the active region at the start of \sunrise{} observations (2013 June 12, 23:39 UT) taken from the HMI and AIA observations.
(a) The line-of-sight (LOS) magnetogram from the \textit{SDO}/HMI saturated at $\pm$500\,G. The solid black box marks the field of view (FOV) of \sunrise{}/IMaX and the dashed white box is the \sunrise{}/SuFI FOV.
(b) \textit{SDO}/AIA 171\,\AA\ channel map showing the coronal emission just below $10^6$\,K aligned with the magnetogram in panel (a).
The AIA data are processed using the multi-scale Gaussian normalization technique.
North is up.\label{fig:hmi_aia}
See Sect.\,\ref{sec:obs}.
}
\end{center}
\end{figure*}

Some new observational evidence challenges this simple picture of a direct magnetic connection from the photosphere to the corona. Recently, \citet{2016ApJ...820L..13W} conducted a study where he related coronal loop-like structures seen with the Atmospheric Imaging Assembly \citep[AIA;][]{2012SoPh..275...17L} to magnetograms from the Helioseismic and Magnetic Imager \citep[HMI;][]{2012SoPh..275..207S}. He showed that there are small coronal loop-shaped features with horizontal sizes of only 5\,Mm that are embedded in the same seemingly unipolar magnetic field region in a plage area. From this \citet{2016ApJ...820L..13W} concluded that there must be small-scale opposite magnetic polarities embedded within the plage region that appears to be unipolar to HMI with its limited spatial resolution of no better than 1\arcsec (corresponding to 725\,km on the Sun at disk center). This is similar to the interpretation given earlier by \cite{2013A&A...556A.104P} for even smaller loop-like features with lengths of only 1\,Mm seen with the High-resolution Coronal Imager~\citep[Hi-C;][]{2014SoPh..289.4393K}. While the spatial resolution of the coronal images of Hi-C is about 0.3\arcsec\ (and thus more than three times better than AIA), the conclusions of \cite{2013A&A...556A.104P} had to rely on the comparably poor resolution of HMI, too. In addition, \citet{2016ApJ...820L..13W} showed one example where a loop-like structure is found at the footpoint of a loop with the general appearance of an inverse Y. So there is evidence for a complex magnetic nature of unipolar-looking regions. This implies a non-trivial mapping of the magnetic field from coronal loops to the photosphere that is more complex than the traditional picture summarized above. 

A complex magnetic mapping from the photosphere to the corona has been suggested a decade ago for the base of the solar wind, and it might be applicable also in the context of (longer) coronal loops. In their study of the initial acceleration of the solar wind \citet{2005Sci...308..519T} found upflows at coronal temperatures, but downflows at transition region temperatures (0.1\,MK) at the base of the magnetic funnel connecting the two temperature regimes. They interpreted this as feeding of the coronal funnels from the side at some height with mass and energy through continuous reconnection. In a numerical model, \citet{2013ApJ...770....6Y} confirmed this scenario, highlighting the role of small bipoles being driven to the (unipolar) footpoints of coronal funnels that are then connected to the solar wind.

All this provides evidence that the mapping of the magnetic field from the corona to the actual footpoints might be not as simple as thought before. We will use high-resolution magnetic data from \sunrise\ \citep{2010ApJ...723L.127S,solanki2016} to investigate the magnetic structure at the footpoints of coronal loops (Sect.\,\ref{sec:obs}). These data provide a spatial resolution about six times better than HMI.

We find that there is an abundance of small-scale mixed polarity at the loop footpoints (Sect.\,\ref{sec:mixed}) that is detected witnessing to the good resolution and sensitivity of \sunrise{}/IMaX. This leads to a scenario in which coronal loops might be mass-loaded and energized by continuous reconnection near their footpoints (Sect.\,\ref{sec:jets}) through the interaction of the major magnetic polarity with small-scale opposite-polarity parasitic magnetic concentrations.

\section{Observations}\label{sec:obs}     

\begin{figure*}[t]
\begin{center}
\includegraphics[width=\textwidth]{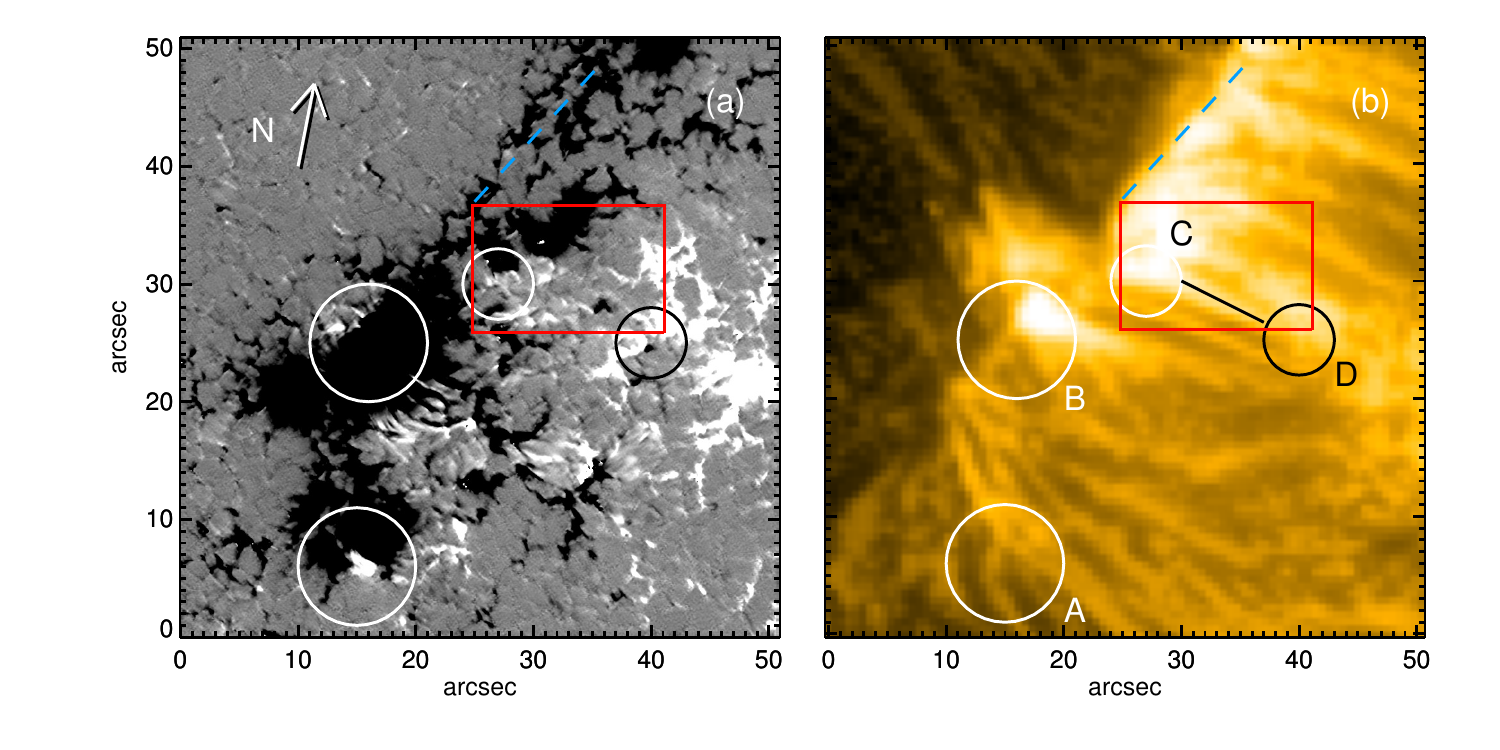}
\caption{Relation of small-scale magnetic field to coronal loops.
Panel (a) shows the \sunrise{}/IMaX LOS magnetogram at 23:45 UT (saturated at $\pm250$\,G; North is indicated by the arrow).
Panel (b) displays the cotemporal 171\,\AA\ channel image of AIA showing emission just below $10^6$\,K at the same time.
The FOV presented here is indicated by the black box in Figure\,\ref{fig:hmi_aia}.
Four regions of interest, marked A--D, are denoted by circles. The circles in the two panels lie at the same location relative to the solar surface.
The black line from circle C to D highlights a short loop that is also seen in the zoom in Figure\,\ref{fig:maps}c.
The red rectangle shows the FOV displayed in Figure\,\ref{fig:maps}.
The blue dashed line marks the footpoint locations of an arcade of coronal loops.
See Sect.\,\ref{sec:mixed}.
\label{fig:imax_aia}}
\end{center}
\end{figure*}

During its second flight, the balloon-borne solar observatory \sunrise{}~\citep{2010ApJ...723L.127S,solanki2016,2011SoPh..268....1B,2011SoPh..268..103B,2011SoPh..268...35G} observed an emerging active region AR 11768, on 2013 June 12 at 23:39 UT, away from disc center at ($280\arcsec,-197\arcsec$), in the southern hemisphere. 
The Imaging Magnetograph eXperiment~\citep[IMaX;][]{2011SoPh..268...57M} onboard \sunrise{} recorded the full Stokes vector for the magnetically sensitive Fe\,{\sc i}\,5250.2\,\AA\ line at eight wavelength positions covering the line and continuum with a cadence of 36.5\,s, for a period of 17\,minutes.
The \sunrise{} Filter Imager~\citep[SuFI;][]{2011SoPh..268...35G} recorded high-cadence ($\approx$7\,s), high-resolution (diffraction limited, 0.07\arcsec\textendash 0.1\arcsec) Ca\,{\sc ii}\,H narrowband filter images at 3968\,\AA\ and wide-band images at 3000\,\AA.
For our study we use the reconstructed IMaX line-of-sight (LOS) magnetic field maps\footnote{For simplicity, in the rest of the text we refer to IMaX LOS magnetic field maps as magnetograms} obtained from SPINOR inversions~\citep{2000A&A...358.1109F} and chromospheric filter images from SuFI~\citep[details of the IMaX and SuFI data reduction can be found in][]{solanki2016}.

Our aim here is to study the details of the magnetic field distribution at the footpoints of coronal loops and their connection to the cooler lower atmosphere.
To this end, we use EUV observations of the solar corona from AIA onboard the \textit{Solar Dynamics Observatory} \citep[\textit{SDO};][]{2012SoPh..275....3P}.
In particular, we focus on the AIA 171\,\AA\ filter showing the emission primarily by Fe\,{\sc ix} at a characteristic temperature of ${\log}\,T$\,[K]${=}5.8$. We supplement these AIA data with LOS  magnetograms from HMI on \textit{SDO}, mostly to compare them to the high-resolution IMaX magnetograms.

To align the data from the different instruments, we use a cross-correlation technique.
In the case of HMI and IMaX we first degrade the spatial scale of IMaX ($0.0545\arcsec\,\text{pixel}^{-1}$) to the HMI pixel scale ($0.5\arcsec\,\text{pixel}^{-1}$) by rebinning the IMaX data. 
Next, the first LOS magnetogram from the IMaX time series is translated and rotated to get the spatial offsets and angle of rotation with maximum correlation with respect to the near simultaneous HMI LOS magnetogram.
The following step is to align IMaX and SuFI data. In order to do this, we first rebin SuFI data (plate scale of $0.02\arcsec\,\text{pixel}^{-1}$) to the plate scale of IMaX. Then, we calculate the shifts between these two instruments by comparing the IMaX 
inverted temperature map at optical depth of unity (at 5000\,\AA) with the SuFI 3000\,\AA\ wide-band image.
The alignment between SuFI and AIA data is achieved first through the visual identification of features (mainly reconnection brightening signatures) common to both datasets.
In a second step, we cross-correlate these features for finer offsets.
The final alignment between all the datasets (i.e. IMaX, SuFI, HMI, and AIA) is accurate to the pixel scale of AIA ($0.6\arcsec\,\text{pixel}^{-1}$).
This corresponds to about 11 pixels in IMaX and re-scaled SuFI data.

The contextual maps of photospheric HMI LOS magnetic flux density and AIA 171\,\AA\ coronal emission at the beginning of the \sunrise{} observations are shown in Figure\,\ref{fig:hmi_aia}.
To enhance the contrast for display purposes, the AIA date are processed using the multi-scale Gaussian normalization technique~\citep{2014SoPh..289.2945M}.
The full field of view (FOV) of the IMaX data is $51\arcsec\times51\arcsec$, while SuFI has a smaller $16\arcsec\times37\arcsec$ FOV (see Figure\,\ref{fig:hmi_aia}a).

IMaX mostly covers the negative (black) magnetic polarity of the active region and, within the FOV common to both IMaX and SuFI, the HMI magnetogram shows the presence of a pore with negative polarity (top part of SuFI FOV near solar $X,Y=285$\arcsec,$-$190\arcsec; see Figure\,\ref{fig:hmi_aia}a). 
We will concentrate on this pore in the discussion in Sect.\,\ref{sec:jets}.
Often such pores are associated with the footpoints of coronal loops, which is confirmed by the identification of loops in the AIA 171\,\AA\ channel at the same location. 
Within the FOV of IMaX the images by AIA in the 171\,\AA\ channel show numerous coronal loops (Figure\,\ref{fig:hmi_aia}b).
Most of these loops are connected to the main large-scale opposite-polarity structures just west (right) of the IMaX FOV as seen in the HMI magnetogram (white patch in Figure\,\ref{fig:hmi_aia}a).

In the rest of the analysis, we keep the orientation of IMaX observations fixed (i.e.\ north points to about 11\textdegree\ in the clockwise direction).
This is to avoid the smoothening and degradation of IMaX data due to rotation and instead rotate the HMI and AIA data accordingly.

\section{Mixed polarities at loop footpoints}\label{sec:mixed}    

\begin{figure*}[t]
\begin{center}
\includegraphics[width=\textwidth]{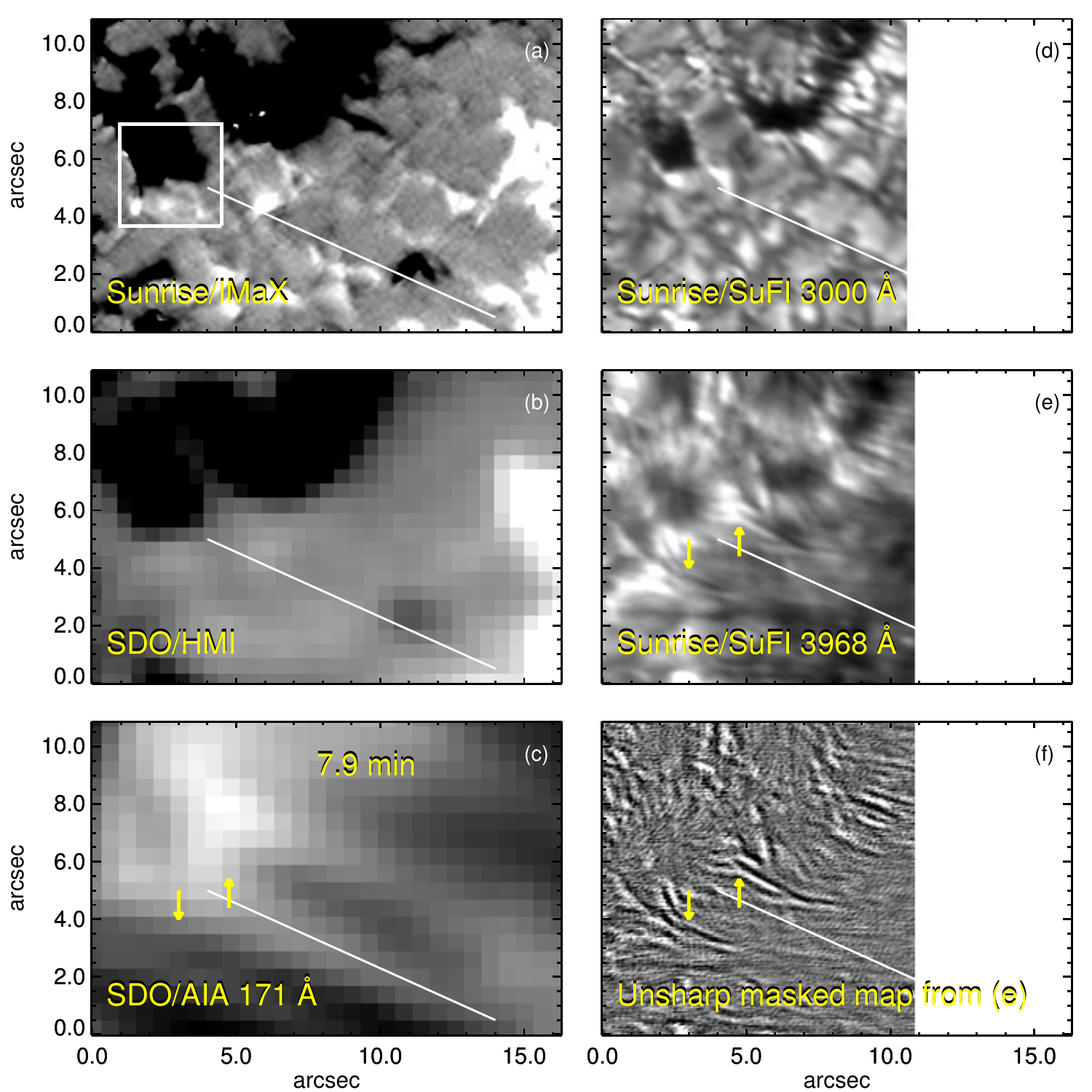}
\caption{Detailed view of the atmosphere including the photospheric magnetic field, the chromosphere, and the corona.
The maps show various observables near region C in Figure\,\ref{fig:imax_aia} within the red rectangle shown there.
These snapshots are taken at 23:47 UT.
(a) \sunrise{}/IMaX LOS magnetogram saturated at $\pm$250\,G. The white box encloses the area used to calculate the magnetic flux from negative and positive polarities (see Figure\,\ref{fig:imax_flux}).
(b) Co-spatial \textit{SDO}/HMI LOS magnetogram saturated at $\pm$250\,G. 
(c) \textit{SDO}/AIA 171\,\AA\ channel image. The time elapsed in minutes since the start of \sunrise{} observations is displayed.
In both the HMI and AIA maps the individual pixels are visible.
(d) \sunrise{}/SuFI wide-band map at 3000\,\AA\ displaying photospheric granulation.
(e) \sunrise{}/SuFI Ca\,{\sc{ii}}\,H narrowband image at 3968\,\AA\ exhibiting lower chromospheric structures.
(f) The Ca\,{\sc{ii}}\,H image from panel (e) after the application of unsharp masking to highlight the small-scale features.
The white solid line in all the panels outlines a coronal loop (see Figure\,\ref{fig:imax_aia}b).
The arrows in panels (c), (e), and (f) point to the location of two inverse Y-shaped jet features.
An animations of this figure is available online.
See Sects.\,\ref{sec:mixed} and \ref{sec:jets}.
\label{fig:maps}}
\end{center}
\end{figure*}

In general, when using magnetic field observations with low or moderate spatial resolution (1\arcsec\ or worse), the footpoint of a coronal loop appears to be rooted in a seemingly unipolar magnetic feature.
In particular, this is the case if it is a strong concentration of unipolar flux, e.g.\ as found in plages or pores.
This is also the case with the HMI data presented here.
A close examination of the locations of loop footpoints identified in AIA maps shows that these are predominantly unipolar magnetic field structures as seen by HMI (see Figure\,\ref{fig:hmi_aia}).

However, what looks unipolar in HMI shows a multitude of small-scale opposite magnetic polarities at higher resolution.
The co-spatial and near simultaneous IMaX magnetograms with about six times better spatial resolution than HMI reveal an abundance of small-scale opposite-polarity magnetic elements close to the dominant polarity.
Actually, in the high-resolution IMaX magnetogram plotted in Figure\,\ref{fig:imax_aia} it seems that the opposite-polarity features (white) appear favorably located at the edges of the concentrations of the main polarity (black).
Zooming into the IMaX magnetogram near a pore (Figure\,\ref{fig:maps}a) highlights the presence of small opposite-polarity features at the edges of the main polarity.
The corresponding HMI image (Figure\,\ref{fig:maps}b) does not unambiguously show these small-scale features.

The main question is how the coronal loops relate to this fine-scale magnetic structure, if they do at all.
Overall the coronal loops connect the two main polarities of the active region (Figure\,\ref{fig:hmi_aia}).
Typically, the IMaX FOV covers only one leg of each loop (Figure\,\ref{fig:imax_aia}), but we can nicely follow that loop by eye down to its footpoint and associate it to a patch of photospheric magnetic field, which we take to be its footpoint.
We highlight four regions of interest A to D, in Figure\,\ref{fig:imax_aia} that show different features of how the coronal loops are related to the magnetic field. By comparing the FOV plotted in Figs.\,\ref{fig:hmi_aia} and \ref{fig:imax_aia}, it is clear that the highlighted regions A to D are footpoint locations of different loops. In the following we describe each region in detail.

(A) Here, a coronal loop seems to end in a bigger pore in the southern part of the IMaX FOV.
Comparing the position of the loop footpoints and the IMaX magnetogram, the footpoint is rooted in the vicinity of the location where one would expect an interaction between the main polarity (black) with the small opposite polarity (white).
Actually, this patch of opposite polarity is so strong and big that it is also slightly visible in HMI (see Figure\,\ref{fig:hmi_aia}), which is not the case for the other opposite-polarity features.

(B) Coronal loops coming from the west (``right'') of the FOV shown in Figure\,\ref{fig:imax_aia} are rooted on the far side of the main polarity.
The loops cross the main polarity all the way to come down in a region where also small opposite magnetic polarities are present.

(C) Shorter loops are also rooted in the near side of the main negative (black) polarity, but again in a region with small-scale opposite polarities.
Actually, the feature (C) shows the footpoint of a short loop where we can identify the other footpoint also in the FOV of IMaX.

(D)~~ At the western (``right'') footpoint of the short loop (indicated by  a black line in Figure\,\ref{fig:imax_aia}b) there is a dominant positive (white) polarity.
Here, too, a small-scale opposite (now black) polarity is present.
In this case we can show that there are minor opposite magnetic polarity elements at both footpoints of the (short) coronal loop connecting regions C and D. 
This is a fortunate case, because due to the limited IMaX FOV we see only one footpoint of most loops with IMaX.

Summarizing these cases we find that mixed polarity field is common at the base of coronal loops.
The time evolution of the IMaX magnetograms shows that the minor small opposite-polarity elements continually interact with the dominant magnetic polarity, which is larger in size.
The interaction is mainly through flux cancellation in time, which may trigger magnetic reconnection that could supply mass and energy into the overlying coronal loop.

\begin{figure}[t]
\begin{center}
\includegraphics[width=0.45\textwidth]{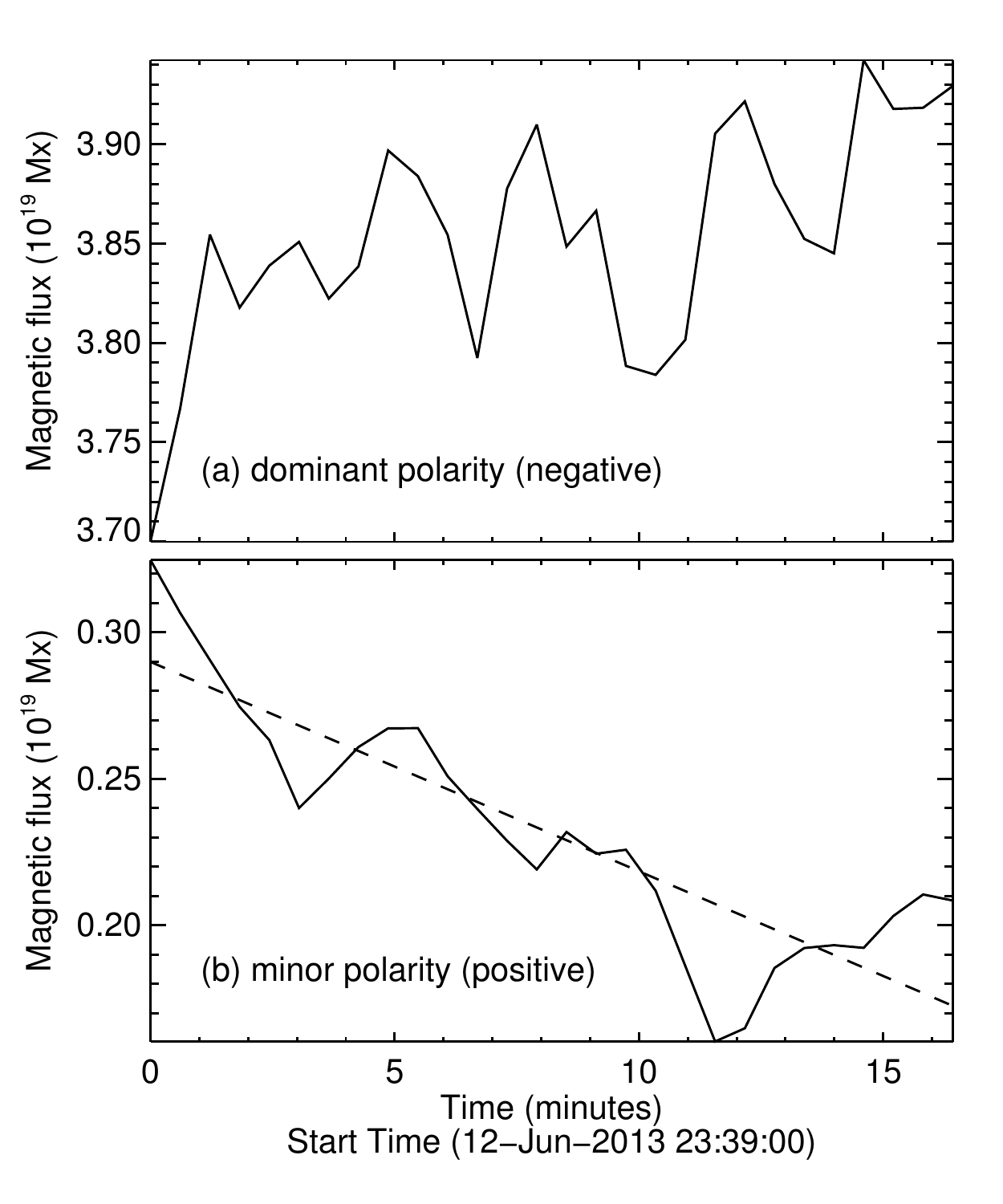}
\caption{Change of magnetic flux at the footpoint of a coronal loop (region C in Figure\,\ref{fig:imax_aia}).
Considering only pixels with magnetic flux densities above 10\,G to avoid noise, we separately plot the flux integrated over the area covered by negative polarity (top panel) and positive polarity (bottom).
The spatial integration is limited to the area outlined by the white box in Figure\,\ref{fig:maps}(a).
The dashed line in the bottom panel shows a linear fit to the observed flux to calculate the rate of change of magnetic flux.
See Sect.\,\ref{sec:mixed}.
\label{fig:imax_flux}}
\end{center}
\end{figure}

To illustrate the interaction of the parasitic opposite polarity with the main polarity we investigate the evolution of the magnetic flux in time.
In Figure\,\ref{fig:imax_flux} we plot the magnetic flux
as a function of time, separately for negative and positive polarity features spatially integrated over the white box marked in Figure\,\ref{fig:maps}a. This represents the change of magnetic flux at the eastern footpoint of the short coronal loop connecting circles C and D in Figure\,\ref{fig:imax_aia} (see also white line in Figure\,\ref{fig:maps}).
For this analysis only those pixels are taken into account where the absolute magnetic flux density is greater than 10\,G (to be above the noise level).

At one footpoint of a short loop in region C we find the negative (black) polarity flux to increase and the positive polarity flux to decrease.
Because this is an emerging active region, it is clear that the flux of main polarity in this part of the active region, i.e.\ the negative one, is increasing.
In contrast, the flux of parasitic small-scale opposite (here positive) polarity decreases over 17\,minutes during the \sunrise\ observations (Figure\,\ref{fig:imax_flux}b). This decrease of magnetic flux of the minor polarity over time is indicative of magnetic flux cancellation.
From the time evolution of magnetic flux of the minor polarity (Figure\,\ref{fig:imax_flux}b), we find that the rate of change of the minor polarity's flux is about  $10^{15}\,\text{Mx\,s}^{-1}$.
The online movie associated with Figure\,\ref{fig:maps} confirms that the parasitic positive polarity features are not moving out of the white box that we use to estimate the flux cancellation rate (Figs.\,\ref{fig:maps}a and \ref{fig:imax_flux}). Instead, they  disappear in situ while moving toward the negative polarity. When increasing the size of the white box the flux cancellation rate would slightly increase, indicating flux cancellation in a larger area. Therefore, the estimated flux cancellation rate within the white box is only a lower limit. 
This flux cancellation indicates the presence of magnetic reconnection between the main and the parasitic opposite polarities that will be further discussed in Sect.\,\ref{sec:jets}.

The zoom into the region C in Figure\,\ref{fig:maps}a--c emphasises that by using the HMI data alone, one would not conclude that the footpoint of the coronal loop in that region is associated with mixed polarities. %
Only the high-resolution IMaX observations reveal this.
Most of the magnetic field sub-structure seen in IMaX is barely distinguishable in the HMI map. 
This further implies that studying the flux evolution at loop footpoints as shown in Figure\,\ref{fig:imax_flux} is not reliable if one would use magnetograms with 1\arcsec\ resolution only.
The mixed polarities at the loop footpoints have consequences also for the chromospheric dynamics and in particular for the mass and energy supply to the corona that is otherwise hidden to the HMI observations.

In addition to the clear cases of coronal loops associated with small-scale mixed polarity field described above (i.e.\ regions A\textendash D), we also see an arcade of densely spaced loops rooted in a mostly negative polarity region (dashed line in Figure\,\ref{fig:imax_aia}). Here, too, we see minor opposite-polarity regions near the loop footpoints, even though they are not as prominent as in the other regions (A\textendash D) and transient in nature with lifetimes of only a few minutes. The intermittency in the parasitic field at the footpoints of this arcade could be due to two reasons: (1) the positive polarity flux is rapidly removed through flux cancellation, e.g.\ as plotted in Figure\,\ref{fig:imax_flux}, and/or (2) the parasitic elements are at the resolution and sensitivity limits of the IMaX instrument. The former reason (1) further substantiates our suggestion that small-scale mixed polarity field is ubiquitous at coronal loop footpoints. The latter option (2) hints that there is a copious amount of magnetic flux hidden even to the current highest resolution magnetic field observations. Both of these possibilities point to an important role of small-scale mixed magnetic field at loop footpoints in the energetics of the solar atmosphere above active regions.

To heat the corona above an active region requires an energy flux of $\approx10^7$\,ergs\,cm$^{-2}$\,s$^{-1}$~\citep[][]{1977ARA&A..15..363W}. To investigate if the flux cancellation studied here can provide this energy, we make an order of magnitude estimate of the average magnetic energy flux, $F_\text{mag}$ (see the appendix for details). From Equations \,\ref{eq:flux} and \ref{eq:flux2}, we estimate that $F_\text{mag} \approx 10^9\,\text{erg}\,\text{cm}^{-2}\,\text{s}^{-1}$. This average energy flux due to small-scale reconnection is significantly larger than the photospheric Poynting flux of $\approx\,5\times 10^7\,\text{erg}\,\text{cm}^{-2}\,\text{s}^{-1}$, generated due to convective motions shuffling the magnetic footpoints in a plage region~\citep{2015PASJ...67...18W}. This suggests that the flux cancellation seen here indeed provides the energy input into the coronal loops associated with the mixed polarity regions. However, what fraction of our estimated average magnetic energy flux actually reaches coronal heights to support the plasma there remains an open question.

\section{Relation to Chromospheric Structures}\label{sec:jets}        

To investigate the mapping of the magnetic structures into the corona we study high-resolution Ca\,{\sc{ii}}\,H images acquired by \sunrise/SuFI.
For this we concentrate on a small region near feature C that is displayed in Figure\,\ref{fig:maps}(e) and (f).
For context, there we also display a photospheric image taken by SuFI in 3000\,\AA\ wide-band channel showing the granulation and two small pores (Figure\,\ref{fig:maps}(d).

At the location of the mixed polarities at the base of the coronal loop in the chromospheric image taken in the core of Ca\,{\sc{ii}}\,H at 3968\,\AA\ (Figure\,\ref{fig:maps}(e)) jet-like features are visible.
They are clearest in the unsharp masked Ca\,{\sc{ii}} map (Figure\,\ref{fig:maps}(f)), and two of them are highlighted by arrows in the figure. 
These jets have a length of 3\,Mm\footnote{The length of the jets might depend on the chromospheric contribution in the spectral region selected by the SuFI filter, i.e.\ by the filter width.} and a lifetime of at least 3\,minutes. 
They appear to be co-spatial with the observed location of the photospheric flux cancellation (see Figure\,\ref{fig:imax_flux}).

The characteristic inverse Y-shape of these jets suggests that they  are the signatures of magnetic reconnection in the solar lower atmosphere.
Their morphology is similar to chromospheric anemone jets observed in Ca\,{\sc ii}\,H movies~\citep{2007Sci...318.1591S} obtained by \textit{Hinode}.
The jets observed here are most probably not related to the rapidly evolving type-II spicules because those have much shorter typical lifetimes of only a few tens of seconds \citep{2007PASJ...59S.655D}. 
However, the association of these jets with longer lived dynamic fibrils or type-I spicules~\citep[lifetimes of 3\textendash 7 minutes;][]{1968SoPh....3..367B,1972ARA&A..10...73B} is not clear.

The observed jets are located at the edges of the coronal loop (see arrows in Figure\,\ref{fig:maps}).
If these jets are physically connected to the loops, they can be the conduits to load plasma into the closed loops through reconnection, which would be similar to the scenario proposed for the origin of the solar wind as discussed in the introduction \citep[see Figure\,5;][]{2005Sci...308..519T}.
Unfortunately, because of the limited resolution of the coronal imaging data of AIA (worse than 1\arcsec) a precise alignment of the chromospheric and the coronal images is not possible.
Most importantly, coronal imaging at this resolution does not provide information if there are also small-scale inverse Y-shaped jets reaching coronal temperatures or if the chromospheric jets are directly related to a possible fine structure of the coronal loop.

Still, the small-scale mixed polarity magnetic field at the footpoints combined with the presence of chromospheric jets at the base of coronal loops suggest a complex mapping of coronal loop to the underlying magnetic field.
Certainly, the magnetic structure at or below the loop footpoint is not as simple as a straight forward magnetic connection as assumed in one-dimensional models.

\begin{figure}[t]
\begin{center}
\includegraphics[width=0.45\textwidth]{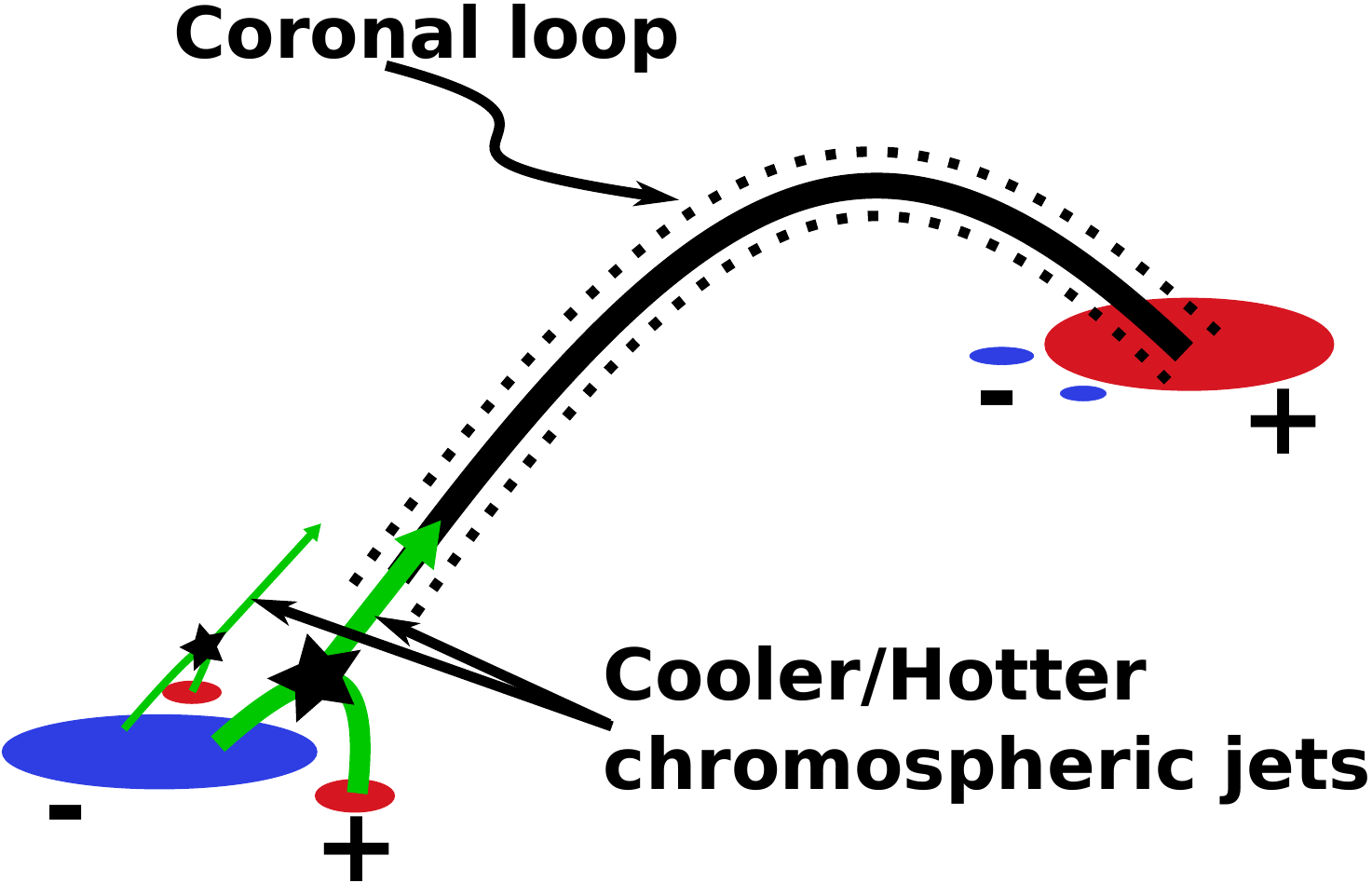}
\caption{Schematic illustration of how a coronal loop might be rooted in the solar lower atmosphere with mixed polarities. The small-scale opposite-polarity magnetic elements interact with the dominant main polarity leading to jet events supplying mass and energy to the coronal loop.
\label{fig:cartoon}}
\end{center}
\end{figure}

\section{Conclusions}\label{sec:concl}   

Our results elucidate that mixed polarity fields can be found at the base of coronal loops where magnetic flux cancellation events are possible. 
The related magnetic reconnection triggered by the cancelling flux may supply energy and (hot) plasma into the solar corona.
Inverse Y-shaped jets are seen in the chromosphere at the locations of flux cancellation, indicating the onset or progress of reconnection.
The lack of coronal observations with a spatial resolution comparable to the \sunrise/SuFI data prevents us from drawing final conclusions on how the coronal features are really connected to the small-scale magnetic structure and reconnection at the footpoints.

These findings might suggest a revision of the traditional picture of the footpoints of coronal loops, which so far have been thought to be unipolar magnetic field structures.
The key findings of our work are the identification of surface mixed polarity magnetic field at the base of coronal loops and the apparent co-spatial jet-like features in the chromosphere. Based on this, we propose that coronal loop footpoints have a sub-structure below the resolution-limit of current telescopes observing the corona ($>$1\arcsec).
This new paradigm is illustrated in Figure\,\ref{fig:cartoon}. 
It is expected that magnetic reconnection leads to plasma jets.
While the cooler jets are visible today in chromospheric imaging, the hotter ones might exist, too, and support the overlying coronal loops.

During one rocket flight, Hi-C showed the potential for high-resolution coronal observations at about 0.3\arcsec\ resolution, and perhaps the second flight will provide data to further test our scenario.
In particular, further efforts of combining high-resolution photospheric magnetic data, e.g.\ during a third flight of \sunrise{}, with coronal observations are required to further constrain our view of the magnetic structure of the footpoints of coronal loops.

With the presence of opposite minor polarities, the magnetic structure at the base of the loop would be more complicated than in the existing models, where even in 3D models there is a direct smooth connection from the photosphere into the coronal loop (see Section\,\ref{sec:intro}). A high-resolution 3D MHD model accounting for the small-scale mixed polarities will be needed to investigate the impact of the mixed polarities at each footpoint. Already the ubiquitous presence of these mixed polarities suggests that they could be of major importance in the heating of the loop and how it is fed with mass.

\begin{acknowledgements}
We thank the anonymous referee for his/her useful comments. L.P.C. acknowledges funding by the Max-Planck Princeton Center for Plasma Physics and funding from the European Union's Horizon 2020 research and innovation programme under the Marie Sk\l{}odowska-Curie grant agreement No.\,707837.
The German contribution to \sunrise{} and its reflight was funded by the Max Planck Foundation, the Strategic Innovations Fund of the President of the Max Planck Society (MPG), DLR, and private donations by supporting members of the Max Planck Society, which is gratefully acknowledged. The Spanish contribution was funded by the Ministerio de Econom\'{i}a y Competitividad under Projects ESP2013-47349-C6 and ESP2014-56169-C6, partially using European FEDER funds. The HAO contribution was partly funded through NASA grant number NNX13AE95G. \textit{SDO} data are the courtesy of NASA/\textit{SDO} and the AIA, and HMI science teams. This work was partly supported by the BK21 plus program through the National Research Foundation (NRF) funded by the Ministry of Education of Korea. The National Solar Observatory (NSO) is operated by the Association of Universities for Research in Astronomy (AURA) Inc. under a cooperative agreement with the National Science Foundation. The National Center for Atmospheric Research is sponsored by the National Science Foundation. This research has made use of NASA's Astrophysics Data System.
\end{acknowledgements}

\appendix
\section{Estimation of the Energy Flux\label{sec:appn}}

We provide an order of magnitude estimate of the energy flux available to the upper atmosphere due to a typical flux cancellation event as discussed at the end of Sect.\,\ref{sec:mixed}. The energy density, $e_{\text{mag}}$, of a magnetic field $\bm{B}$ is given by
\begin{equation}
e_\text{mag}=\frac{1}{2\mu_0}B^2\label{eq:emag},
\end{equation}
where $\mu_0$ is the magnetic permeability and $B$ is the magnetic flux density. The rate of change of the energy density, $\dot{e}_\text{mag},$ over a time-scale $\tau$ is simply
\begin{equation}
 \dot{e}_\text{mag} = \frac{e_\text{mag}}{\tau}\label{eq:emagdot}. 
 \end{equation}

Let $A$ be the cross-sectional area of a magnetic flux element whose flux, $\Phi$, is disappearing over $\tau$. Then the rate of change of magnetic flux, $\dot{\Phi}$, of that element can be written as
 \begin{equation}
\dot{\Phi}=\frac{BA}{\tau}.\label{eq:phidot}
 \end{equation}
Inserting the above equation in Eq.\,\ref{eq:emagdot} yields 
\begin{equation}
 \dot{e}_\text{mag} = \frac{1}{2\mu_0}\frac{B\dot{\Phi}}{A}\label{eq:emagdo2t}.
\end{equation}

Here, we assume that the energy released during the observed flux cancellation is fed at the base of the loop over a heating length scale, $L_\text{H}$. The average magnetic energy flux, $F_\text{mag}= \dot{e}_\text{mag}L_{H}$, available at the coronal loop footpoint is then
\begin{equation}
F_\text{mag} = \frac{1}{2\mu_0}\frac{B\dot{\Phi}L_\text{H}}{A}\label{eq:flux}.
\end{equation}
Both 3D MHD models~\citep[e.g.][]{2011A&A...530A.112B} and nonlinear force-free field models~\citep[e.g.][]{2014ApJ...793..112C} show that the average heating rate drops by 4\textendash 6 orders of magnitude 
from the photosphere to the base of corona at 4 Mm, which yields $L_\text{H}\approx 500$\,km. The heating length-scale can be physically understood in terms of decaying magnetic energy above certain height, which depends on the closing down of magnetic loops below that height. Using the same IMaX data, \citet{requerey2016} analyzed region A (see Figure\,\ref{fig:imax_aia}) and estimated that closed magnetic field lines between the minor positive and dominant negative polarity there reach heights of about 500\,km. This supports our choice for $L_\text{H}$.

In our observations, the cancelling magnetic features have a size of about 500\,km ($A=25\times 10^4\,\text{km}^2$), corresponding to a few times of a typical magnetic element. With a magnetic field strength on the order of 1\, kG, the observed flux cancellation rate of $\dot{\Phi}=10^{15}\,\text{Mx s}^{-1}$ (Sect.\,\ref{sec:mixed}), and $L_\text{H} \approx 500\,\text{km}$, we obtain
\begin{equation}
F_\text{mag} \approx 10^9\,\text{erg}\,\text{cm}^{-2}\,\text{s}^{-1}.\label{eq:flux2}
\end{equation}
This average energy flux is two orders of magnitude larger than the typical vertical Poynting flux in a plage region as found by \citet{2015PASJ...67...18W}. Therefore, we suggest that the reconnection at coronal loop footpoints might provide required energy input to balance the coronal energy losses.

\end{document}